\documentclass[onecolumn]{emulateapj}  
\usepackage{epstopdf}
\usepackage{ulem}
\usepackage{amssymb}
\usepackage{natbib}
\usepackage{times}
\usepackage{graphicx}
\usepackage[usenames,dvipsnames]{pstricks}

\usepackage{psfrag}
\usepackage[colorlinks=true,linkcolor=blue,citecolor=blue]{hyperref}
\def\araa{ARA\&A}
\def\apj{ApJ}
\def\apjl{ApJ}
\def\apss{Ap\&SS}
\def\aap{A\&A}
\def\jcap{J. Cosmology Astropart. Phys.}
\def\prd{Phys.~Rev.~D}
\def\pasj{PASJ}
\def\ssr{Space~Sci.~Rev.}
\def\nat{Nature}
\def\physrep{Phys.~Rep.}

\newcommand{\be}{\begin{equation}}
\newcommand{\ee}{\end{equation}}
\newcommand{\bary}{\begin{eqnarray}}
\newcommand{\eary}{\end{eqnarray}}
\newcommand{\en}{E_\nu}

\shorttitle{TeV $\gamma$-ray, neutrinos and ultra-high-energy cosmic ray fluxes in Mrk 421}
\shortauthors{Fraija N.}
\begin{document}
\title{TeV $\gamma$-ray fluxes  from the long campaigns on Mrk421 as constraints\\ on the emission of  TeV-PeV Neutrinos and UHECRs}
\author{N. Fraija\altaffilmark{1} and A. Marinelli \altaffilmark{2}} 
\email{nifraija@astro.unam.mx and antonio.marinelli@pi.infn.it} 
\altaffiltext{1}{Luc Binette-Fundaci\'on UNAM Fellow. Instituto de Astronom\' ia, Universidad Nacional Aut\'onoma de M\'exico, Circuito Exterior, C.U., A. Postal 70-264, 04510 M\'exico D.F., M\'exico.}
\altaffiltext{2}{I.N.F.N. \& Physics Institute Polo Fibonacci Largo B. Pontecorvo, 3 - 56127 Pisa, Italy.}
\date{\today} 
\begin{abstract}
Long TeV $\gamma$-ray campaigns have been carried out to study the spectrum, variability and duty cycle of the  BL Lac object Markarian 421. These campaigns  have given some evidence of the presence of protons in the jet: i)  Its spectral energy distribution which shows two main peaks; one at low energies ($\sim$ 1 keV) and the other at high energies  (hundreds of GeV), has been described by using synchrotron proton blazar model. ii) The  study of the variability at GeV $\gamma$-rays and X-rays has indicated  no significant correlation. iii)  TeV $\gamma$-ray detections without activity in X-rays, called "orphan flares" have been observed in this object.\\ 
Recently,  The Telescope Array Collaboration reported the arrival of 72 ultra-high-energy cosmic rays  with some of them possibly related to the direction of Markarian 421. The IceCube Collaboration reported the detection of 37 extraterrestrial neutrinos in the TeV - PeV energy range collected during three consecutive years. In particular, no neutrino track events were associated with this source. In this paper,  we consider the proton-photon interactions to correlate the TeV $\gamma$-ray fluxes reported by long campaigns with the neutrino and ultra-high-energy cosmic ray observations around this blazar.  Considering the results reported by The IceCube and Telescope Array Collaborations, we found  that only from $\sim$ 25\% to 70\% of TeV fluxes described with a power law function with exponential cutoff can come from the proton-photon  interactions.
%
\end{abstract}


\section{Introduction}
At a distance of 134.1 Mpc (z=0.03), the BL Lac object  Markarian 421 (Mrk 421) \citep{1991rc3..book.....D, 2005ApJ...635..173S} is one of the closest known and brightest sources in the extragalactic X-ray/TeV sky. In X-rays, this object has been observed for more than 14 years,  measuring fluxes ranging from few to hundreds mCrab \citep{2010PASJ...62L..55I,2009ApJ...691L..13D, 2012ApJ...759...84N, 2009ATel.2292....1K, 2008ATel.1574....1C,2006ATel..848....1L,2013ATel.4974....1B}. In particular the All-Sky Monitor (ASM) on board of the Rossi X-ray Timing Explorer (RXTE) was monitoring the X-ray sky from 1996 to 2011. This instrument observed continually the sky in four energy bands (1.5-3, 3-5, 5-12 and 1.5-12 keV) \citep{1996ApJ...469L..33L}.  The maximum X-ray flux measured was  $\sim$ 4.3 counts s$^{-1}$ which corresponds to $\sim$ 55.65 mCrab. Furthermore,  \citet{2009A&A...502..499R} through RXTE/ASM data  estimated the duty cycle of this source, finding  a constant value of 18.1\%  for a standard deviation of  3$\sigma$.   In TeV energies, simultaneous observations have been carried out  by different telescopes based on Imaging Atmospheric Cherenkov Techniques (IACT) \citep{1992Natur.358..477P,1996Natur.383..319G,2011ApJ...738...25A, 2002ApJ...575L...9K, 2007ApJ...663..125A, 2005A&A...437...95A, 2002A&A...393...89A, 2003A&A...410..813A,2007ApJ...662..199C,2014APh....54....1A} and air shower arrays (ASAs) \citep{2011ApJ...734..110B,2014ApJ...782..110A,2003ApJ...598..242A}.   In particular, the Milagro experiment observed this source consecutively for a period of 906 days with a significance of 7.1 standard deviations. During these almost 3 years of observations, Milagro measured an average flux for energies above 1 TeV  equal to $(0.205\pm0.030)\times 10^{-10}\, {\rm cm^{-2}\,s^{-1}}$ for a spectral index of 2.3. With this observatory it was possible to estimate also the duty cycle for energies above 1 TeV for different baseline fluxes.  At 3 standard deviations,   the values of duty cycle reported were  $\sim$ 40\% and $\sim$ 27\% for 0.1 Crab and 0.33 Crab of luminosity, respectively \citep{2014ApJ...782..110A}.    For a period of 1.5 years (from 2008 August 5 to 2010 March 12), Mrk 421 was monitored by the Large Area Telescope (LAT) on board of the Fermi satellite. The $\gamma$-ray flux collected above 0.3 GeV  was described by a power law with a photon index of $1.78\pm 0.02$ and average flux of $(7.23\pm 0.16)\times 10^{-8}\, {\rm ph\, cm^{-2}\, s^{-1}}$ \citep{2010ApJ...719.1433A}.  The broadband spectral energy distribution (SED) with two peaks, one of low energy ($\sim$ 0.5 keV), and the other of high energy (at hundreds of GeV) was described through leptonic and hadronic models.  In the leptonic scenario, a one-zone synchrotron self-Compton (SSC) with three accelerated electron populations (through diffusive relativistic shocks with a randomly oriented magnetic field)  has been used \citep{2008ApJ...686..181F, 2010ApJ...719.1433A}.  In the hadronic scenario, the peak at low energies is explained by electron synchrotron radiation whereas the high-energy peak is explained evoking the Synchrotron-Proton Blazar (SPB) model \citep{2001APh....15..121M,2003APh....18..593M}.  In addition, \citet{2010ApJ...719.1433A} investigated the X-ray/GeV $\gamma$-ray correlation using  RXTE/ASM  and  Fermi-LAT data, respectively.  For RXTE/ASM, an energy range of  2-10 keV was used whereas for Fermi-LAT data two energy ranges  0.2 -2 GeV and 2 - 300 GeV were analyzed.  They  observed that the variability in the Fermi fluxes above 2 GeV and below 2 GeV is shorter than the  variability in X-rays,  and they did not find any significant correlation with X-rays in either of the two Fermi energy ranges. It is worth noting that  TeV $\gamma$-ray and X-ray fluxes are usually correlated and  interpreted through the standard one-zone SSC model \citep{1995ApJ...449L..99M, 2008ApJ...677..906F, 2009ApJ...695..596H},  supporting a leptonic scenario.   On the other hand,  two unusual TeV flares  without  X-ray counterpart have happened  \citep{2005ApJ...630..130B,2011ApJ...738...25A}.  Firstly, at around  MJD 53033.4 a TeV flare was observed when the X-ray flux was low. This X-ray flux seems to have peaked 1.5 days before the  $\gamma$-ray flux, and secondly,  the higher TeV flux during the night of MJD 54589.21 was not accompanied by simultaneous X-ray activity.  The last atypical event was interpreted in the framework of hadronic scenario \citep{2015arXiv150104165F}. \\
From the above considerations, Mrk 421 may have the potential to emit high-energy neutrinos and to accelerate protons up to ultra-high energies (UHEs). Recently, The IceCube Collaboration reported the detection of 37 extraterrestrial neutrinos at 5.7$\sigma$  above 30 TeV  \citep{2013arXiv1311.5238I, 2013arXiv1304.5356I, 2014arXiv1405.5303A}. The reconstructed neutrino events in the TeV - PeV energy range have been obtained during three consecutive years of data taking (2010 to 2013).  In particular, no well-located astrophysical neutrino candidates (those with muon tracks in the detector) have been associated with the location of Mrk 421. On the other hand, The Telescope Array  (TA) Collaboration reported a cluster of 72 ultra-high-energy cosmic rays (UHECRs) with energies above 57 EeV.  The cluster of events was centered at R. A.=146$^\circ$.7, Dec.=43$^\circ.2$ and had a Li-Ma statistical significance of 5.1$\sigma$ within 5 years of operation. The field of view of this observatory covers the sky region above -10$^\circ$ of declination having a good sensitivity in the direction of Mrk 421 \citep{2014arXiv1404.5890T}.  In fact, by considering the error reported for the reconstructed directions and the deviation due to extragalactic and galactic magnetic fields, a subset of few events might be linked to the position of Mrk 421.\\
Taking into account the distance of Mrk 421, only  23\% of the UHE proton fraction can come from this BL Lac \citep{2011ARA&A..49..119K}. Then,  our starting point is to plot in a sky-map the neutrino track events detected by The IceCube Collaboration, the 72 UHECRs collected by TA experiment and the BL Lac Mrk 421.   As shown in Figure 1,  a circular region  of 5$^\circ$ around Mrk 421 encloses one UHECR and no neutrino track events. In this work,  we assume the presence of relativistic protons accelerated in the jet and introduce a hadronic model to link the TeV $\gamma$-ray fluxes reported by the long campaigns of the Very Energetic Radiation Imaging Telescope Array System  (VERITAS, \citet{2011ApJ...738...25A}), Milagro \citep{2014ApJ...782..110A} and Whipple \citep{2014APh....54....1A} experiments with at least one UHECR observed in this region by  TA experiment \citep{2014arXiv1404.5890T}.  Additionally, we link these TeV $\gamma$-ray fluxes with the absence of neutrino tracks at the TeV - PeV energy range  reported by The IceCube Collaboration \citep{2013arXiv1311.5238I, 2014arXiv1405.5303A}.  In this work we require the parameters derived from the hadronic model to be able to describe reasonably well the broadband SED of this blazar \citep{2011ApJ...736..131A}.  The paper is arranged as follows. In Section 2 we show a hadronic model  that relates the TeV $\gamma$-ray, HE neutrinos  and UHE proton fluxes through p$\gamma$ interactions. In section 3 we obtain, through Monte Carlo simulation, the neutrino event rate produced by hadronic interactions and detected in a km$^{3}$ neutrino telescope on Earth. In Section 4 we discuss the capability of Mrk 421 to accelerate the UHECRs and estimate the number of expected UHECRs in the TA experiment.    In Section 5 we present a full analysis  and also a discussion of  our results; a brief summary is given in section 6.  We hereafter use primes (unprimes) to define the quantities in a comoving (observer) frame,  k=$\hbar$=c=1 in natural units,  and z=0.03$\simeq$ 0.
 \section{Hadronic  model}
As has been pointed out by \citet{2001APh....15..121M}, relativistic protons in the jet  suggest that  hadronic interactions must be taken into account in describing the broadband SED, as well as neutrino emission. Relativistic protons accelerated at the emitting region and described through a power law by
\be\label{dN_p}
 \frac{dN_p}{dE_p}=A_p \left(\frac{E_p}{{\rm E_0}}\right)^{-\alpha}, 
\ee
can interact with the target photon density given by
\be
n_{\gamma}\simeq\frac{d^2_z}{r^2_d\,\epsilon_{\gamma, {\rm pk}}} (\nu F_\nu)\,,
\label{den}
\ee
where  $\nu F_\nu\simeq L_\gamma/(4\pi d^2_z)$ is the photon flux, $r_d$ is the radius of the emitting region, $\epsilon_{\gamma, {\rm pk}}$ is the peak energy of  target photons and $E_0$ is the normalization proton energy. The  charged ($\pi^+$) and neutral ($\pi^0$) pion production channels  are  $n\,+\,  \pi^{+}$  and $p\,+\,\pi^{0}$, respectively. Subsequently, neutral pions decay into photons  $\pi_{0}\rightarrow\gamma\gamma$, and  charged pions into electrons/positrons and neutrinos $\pi^{\pm}\rightarrow \mu^{\pm}+ \nu_{\mu}/\bar{\nu}_{\mu} \rightarrow e^{\pm}+\nu_{\mu}/\bar{\nu}_{\mu}+\bar{\nu}_{\mu}/\nu_{\mu}+\nu_{e}/\bar{\nu}_{e}$.
\subsection{$\pi^0$ decay products}
From p$\gamma$ interactions, photo-pions and neutrinos typically carry $10\% (\xi_{\pi^0}/2\simeq  0.10)$ and 5\% of the proton's energy $E_p$, respectively.  The energy loss rate due to pion production can be written through the pion cooling time \citep{PhysRevLett.21.1016, PhysRevLett.78.2292}
\begin{equation}
{t'_{\pi^0}}^{-1}=\frac{1}{2\,\gamma^2_p}\int\,d\epsilon\,\sigma_\pi(\epsilon)\,\xi_{\pi^0}\,\epsilon\int dx\, x^{-2}\, \frac{dn_\gamma}{d\epsilon_\gamma} (\epsilon_\gamma=x)\,,
\end{equation}
where $\gamma_p$ is the proton Lorentz factor and $\sigma_\pi(\epsilon)=\sigma_{peak}\approx 9\times\,10^{-28}$ cm$^2$ is the cross section of pion production. Comparing the pion cooling and the dynamical time scale (photo pion efficiency),  we get 
\bary
f_{\pi^0}=\frac{t'_d}{t'_{\pi^0}} \simeq \frac{3\,L_\gamma\,\sigma_{peak}\, \Delta \epsilon_{peak}\, \xi_{\pi^0}}{4\pi\,\Gamma^2\,r_d\,\epsilon_{peak}\,\epsilon_{\gamma,{\rm pk}}}
\cases{
 \frac{\epsilon_{\pi^0,\gamma}}{\epsilon_{\pi^0,\gamma,c}}      &  $\epsilon_{\pi^0,\gamma} < \epsilon_{\pi^0,\gamma,c}$\cr
1                                                                                      &   $\epsilon_{\pi^0,\gamma,c} < \epsilon_{\pi^0,\gamma}$\,,\cr
}
\eary
where $\Delta\epsilon_{peak}\simeq$ 0.22 GeV and $\epsilon_{peak}\simeq$ 0.3 GeV.  Then, we can estimate the relation between the photo-pion and Fermi-accelerated proton fluxes as
 \be\label{esp_f_p}
f_{\pi^0}\,E_p\,\biggl(\frac{dN}{dE}\biggr)_p\,dE_p=\epsilon_{\pi^0,\gamma}\,\biggl(\frac{dN}{d\epsilon}\biggr)_{\pi^0,\gamma}\,d\epsilon_{\pi^0,\gamma}\,.
\ee
Here $\Gamma$ is the bulk  Lorentz factor and {\small $\epsilon_{\pi^0,\gamma,c}$} is the break photo-pion energy.   Taking into account the proton spectrum (eq. \ref{dN_p}) and  the energy fraction carried by photons, we can write the photo-pion spectrum as 
\bary\label{pgammam}
\left(\epsilon^2\,\frac{dN}{d\epsilon}\right)_{\pi^0,\gamma} =A_{p\gamma}\,\epsilon_0^2\,
\cases{
\left(\frac{\epsilon_{\pi^0,\gamma,c}}{\epsilon_{0}}\right)^{-1} \left(\frac{\epsilon_{\pi^0,\gamma}}{\epsilon_{0}}\right)^{-\alpha+3}          & $\epsilon_{\pi^0,\gamma} < \epsilon_{\pi^0,\gamma,c}$\cr
\left(\frac{\epsilon_{\pi^0,\gamma}}{\epsilon_{0}}\right)^{-\alpha+2}                                                                                       &   $\epsilon_{\pi^0,\gamma,c} < \epsilon_{\pi^0,\gamma}$\,.\cr
}
\eary
\noindent Here $A_{p\gamma}=b\,A_\gamma$ is obtained through the flux normalization ($A_\gamma$) of each TeV $\gamma$-ray campaign and the parameter ($0\leq\, b\, \leq 1$) associated to the flux generated by  p$\gamma$ interactions.  The TeV $\gamma$-ray flux  is used to normalize the proton spectrum as
\be\label{Apg}
A_p\simeq f_{p,\gamma} \,A_{p\gamma} =    b\, f_{p,\gamma} \,A_\gamma\,,
\ee
\noindent where $f_{p,\gamma}$ is given by 
\be\label{f_pg}
f_{p,\gamma}= \frac{\Gamma^2\,\epsilon_{peak}\left(\frac{2}{\xi_{\pi^0}}\right)^{\alpha-1}}{6\,r_d\,\eta_\gamma\,\sigma_{peak}\,\Delta\epsilon_{peak}}\,,
\ee
with $\epsilon_0$ the normalization photon energy and $n_{\gamma}$ given by eq. (\ref{den}).
\section{High-energy Neutrino flux}
High-energy photons and neutrinos can be correlated through p$\gamma$ interactions (Fermi-accelerated protons with keV photon targets) \citep[see, e.g.] [and reference therein]{2007Ap&SS.309..407H} as follows
\be\label{neu-pho}
\int \frac{dN_{\nu}}{d\en}\,\en\,d\en=k_\nu\int \left( \frac{dN}{d\epsilon}\,\epsilon\,d\epsilon\right)_{\pi^0,\gamma}\,,
\ee
where $k_\nu$ is 1/4. The integral term on the right represents the TeV $\gamma$-ray flux described by eq.(\ref{pgammam}) and the term on the left is the neutrino flux which can be described as a simple power law
\be
\frac{dN_{\nu}}{d\en}=A_{\nu} \, \left(\frac{\en}{\epsilon_0}\right)^{-\alpha_{\nu}}.
\label{espneu1}
\ee
Here we assume that spectral indices for neutrino and $\gamma$-ray spectra are similar  $\alpha\simeq \alpha_\nu$ \citep{2008PhR...458..173B}. Solving the integrals in eq.  (\ref{neu-pho}),  we found that the normalization factors, for neutrino and photon spectrum, are related by
\be\label{Av_Ag}
A_\nu\simeq 2^{-\alpha}\,b\,A_\gamma\,.
\ee
With these considerations, we can write the expected number of reconstructed neutrino events  in a hypothetical neutrino telescope ($> \en^{th}$) as
\be
N_{ev}=T \rho_{w,i}\,N_A\, V_{eff}\int_{\en^{th}} \sigma_{\nu N}  \frac{dN_\nu}{d\en}\,d\en\,,
\label{evtrate}
\ee
where $T$ is the observation time,   N$_A$=6.022$\times$ 10$^{23}$ g$^{-1}$ is the Avogadro number, $\rho_{w,i}$ is the density of the water/ice, $\sigma_{\nu N}=6.78\times 10^{-35}{\rm cm^2}(\en/TeV)^{0.363}$  is  the neutrino-nucleon cross section \citep{1998PhRvD..58i3009G} and $V_{eff}$ is the $\nu_\mu+\bar{\nu}_\mu$ effective volume, obtained through Monte Carlo simulation for a point source emission at the position of Mrk 421.   For a complete analysis, atmospheric and diffuse neutrino ``backgrounds'' should also be introduced. The cosmic diffuse neutrino flux  is obtained  with the upper limit given as $E^{2}_{\nu}\,d\Phi/dE_{\nu}<2.0\times 10^{-8}\, {\rm GeV\,cm^{-2}s^{-1}sr^{-1}}$  \citep{2001PhRvD..64b3002B} and the atmospheric neutrino flux is described by the Bartol model \citep{2006PhRvD..74i4009B, 2004PhRvD..70b3006B}.
\section{Production of UHE cosmic rays}
Astrophysical objects that accelerate cosmic rays up to UHEs, also might produce TeV $\gamma$-ray and neutrino fluxes through hadronic interactions.  In our model we assume that the accelerated proton spectrum extends up to $\sim 10^{20}$ eV energies \citep{2009NJPh...11f5017K, 2009NJPh...11f5016D, 2003APh....19..559P} and  based on this assumption, we  calculate the  number of events expected in the TA experiment.
\subsection{Hillas condition and deflexions} 
Cosmic rays can be accelerated up to UHEs depending on  both the size ($R$) and the strength of the magnetic field ($\mathcal B$) in the acceleration region. Therefore, the maximum  energy required is $E_{max}=Ze\,\mathcal B\,R\,\Gamma$ with Z  the atomic number \citep{1984ARA&A..22..425H}.  Although the Hillas criterion is a necessary condition and acceleration of UHECRs in AGN jets \citep{2012ApJ...749...63M,2012ApJ...745..196R,2010ApJ...719..459J},  it is far from trivial (see e.g., Lemoine \& Waxman 2009 for a more detailed energetics limit \citep{2009JCAP...11..009L}).  Additional limitations in this process could be mainly caused by  the radiative losses or available time when particles go through the magnetized region. Close to the black hole (BH), the acceleration region is limited by the emitting region ($r_d$) and the strength of the magnetic field (B) in it.  Then, the maximum energy achieved is \citep{2010ApJ...719.1433A,  2012PhRvD..85d3012S,2014ApJ...783...44F}
\be\label{sregion}
E_{max}=Ze\,B\, r_{d}\,\Gamma\,.
\ee
It is worth noting that UHECRs traveling from source to Earth are randomly deviated by galactic (B$_G$) and extragalactic (B$_{EG}$) magnetic fields. By considering a quasi-constant and homogeneous magnetic field,  the deflection angle due to the B$_G$ is
\be\label{thet_G}
\theta_G\simeq 3.8^{\circ}\left(\frac{E_{p,th}}{57 EeV}\right)^{-1} \int^{L_G}_0  | \frac{dl}{{\rm kpc}}\times \frac{B_G}{4\,{\rm \mu G}} |\,,
\ee
where L$_G$ corresponds to the distance of our Galaxy (20 kpc) and due to B$_{EG}$,  the deflection angle can be written as 
\be\label{thet_EG}
\theta_{EG}\simeq 5^{\circ}\left(\frac{E_{p,th}}{57 EeV}\right)^{-1} \left( \frac{B_{EG}}{1.25\,{\rm nG}} \right)\,\left(\frac{L_{EG}}{100\, {\rm Mpc}}\right)^{1/2}\,\left(\frac{l_c}{1\, {\rm Mpc}}\right)^{1/2}\,,
\ee
where $l_c$ is the coherence length \citep{1997ApJ...479..290S,2009JCAP...08..005M}  and $E_{p,th}=57\,{\rm EeV}$ is the energy threshold of TA experiment.  Due to the strength of extragalactic ($B_{EG}\simeq$ 1 nG) and galactic ($B_{G}\simeq$ 4 $\mu$G) magnetic fields,   UHECRs are deflected ($\psi_{EG}\simeq 5^{\circ}$ and $\psi_G\simeq 3.8^{\circ}$) between the true direction to the source, and the observed arrival direction, respectively.  Evaluation of these deflection angles relates the transient UHECR sources with the  high-energy neutrino and $\gamma$-ray emission.

\subsection{Expected Number of  UHECRs} 
The TA experiment, located at 1,400 m above sea level in Millard County (Utah),  is made of  three fluorescence detector (FD) stations and a scintillator surface detector (SD) array \citep{2012NIMPA.689...87A}. It was designed to observe cosmic rays that induce extensive air showers with energies above 1 EeV. This array has an area of $\sim$ 700 km$^2$ and has been in operation since 2008.   To estimate the number of UHECRs,  we take into account the TA  exposure, which  for a point source is given by $\Xi\,t_{op}\, \omega(\delta_s)/\Omega$, where $\Xi\,t_{op}=(5)\,7\times10^2\,\rm km^2\,yr$, $t_{op} $ is the total operational time (from 2008 May 11 to 2013 May 4),  $\omega(\delta_s)$ is an exposure correction factor for the declination of Mrk 421 \citep{2001APh....14..271S} and $\Omega\simeq\pi$.     The expected number of UHECRs above an energy $E_{p,th}$ yields
\be
N_{\tiny UHECR}= F_r\, ({\rm  TA\, Expos.})\times \,N_p, 
\label{num}
\ee
where F$_r$ is the fraction of propagating cosmic rays that survives over a distance $>$ D$_z$ \citep{2011ARA&A..49..119K} and  $N_p$ is calculated from the proton spectrum (eq. \ref{dN_p}) extended to energies higher than $E_{p,th}$.   The expected number can be written as
\bary
N_{\tiny UHECR}=F_r\,\frac{\Xi\,t_{op}\, \omega(\delta_s)}{\Omega\,(\alpha-1)}\,b\, f_{p,\gamma}\,A_\gamma\,{\rm E_0}\,\left(\frac{E_{p,th}}{\,{\rm E_0}}\right)^{-\alpha+1} ,
\label{nUHE1}
\eary
with $f_{p,\gamma}$ given by eq. (\ref{f_pg}).
In addition,  from eqs. (\ref{dN_p}) and (\ref{Apg}) we can compute that the proton luminosity $L_p\simeq  4\pi D^2_z\,F_r E^2_p \frac{dN_p}{dE_p}$ can be written as
\be\label{lum}
L_p= 4\pi D^2_z\,F_r\,b\,f_{p,\gamma}\,A_{\gamma}\,{\rm E_0}^2\,\biggl(\frac{E_p}{{\rm E_0}}\biggr)^{2-\alpha} \,.
\ee
It is worth mentioning that if $E_p=E_{p,th}$,  then  $L_p=L_{UHECR}$. 
\section{Analysis and Results}
\begin{figure}
\vspace{0.4cm}
{\centering
\resizebox*{1.2\textwidth}{0.33\textheight}
{\includegraphics{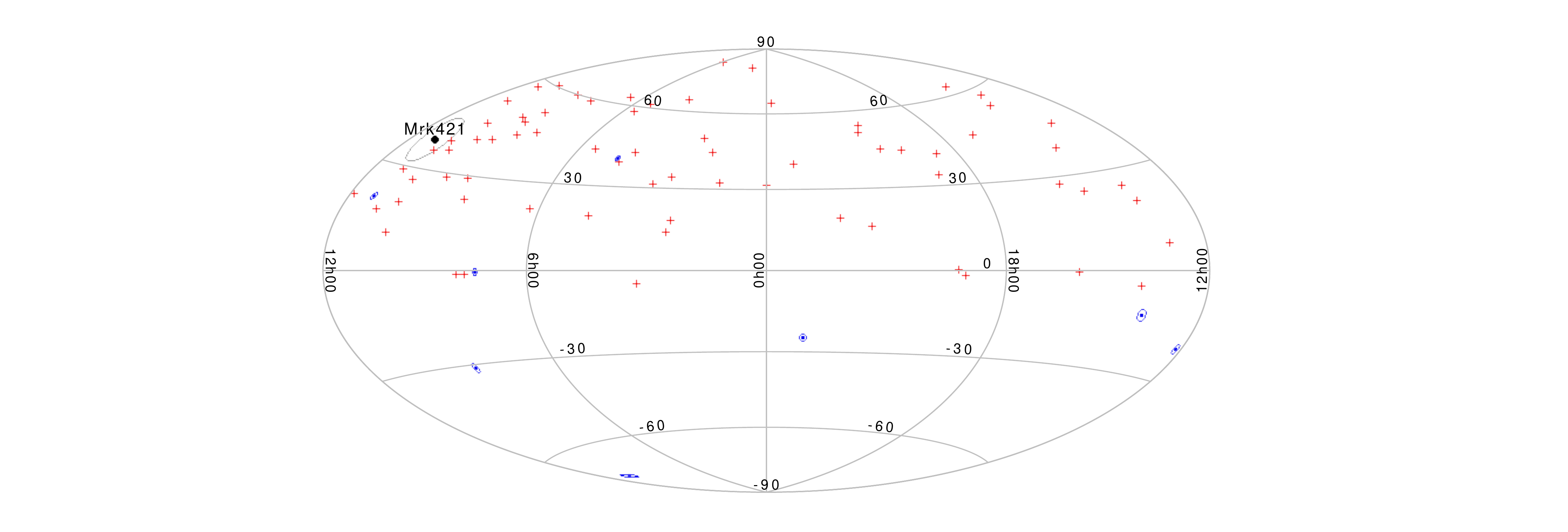}}
}
\caption{Sky-map with the 8 VHE neutrino track events detected by IceCube (blue circles), 72 UHECRs collected by TA experiments (red crosses) and the BL Lac Mrk 421 (black point).}
\label{fit}
\end{figure}

\begin{figure*}
\centering
\includegraphics[width=\textwidth]{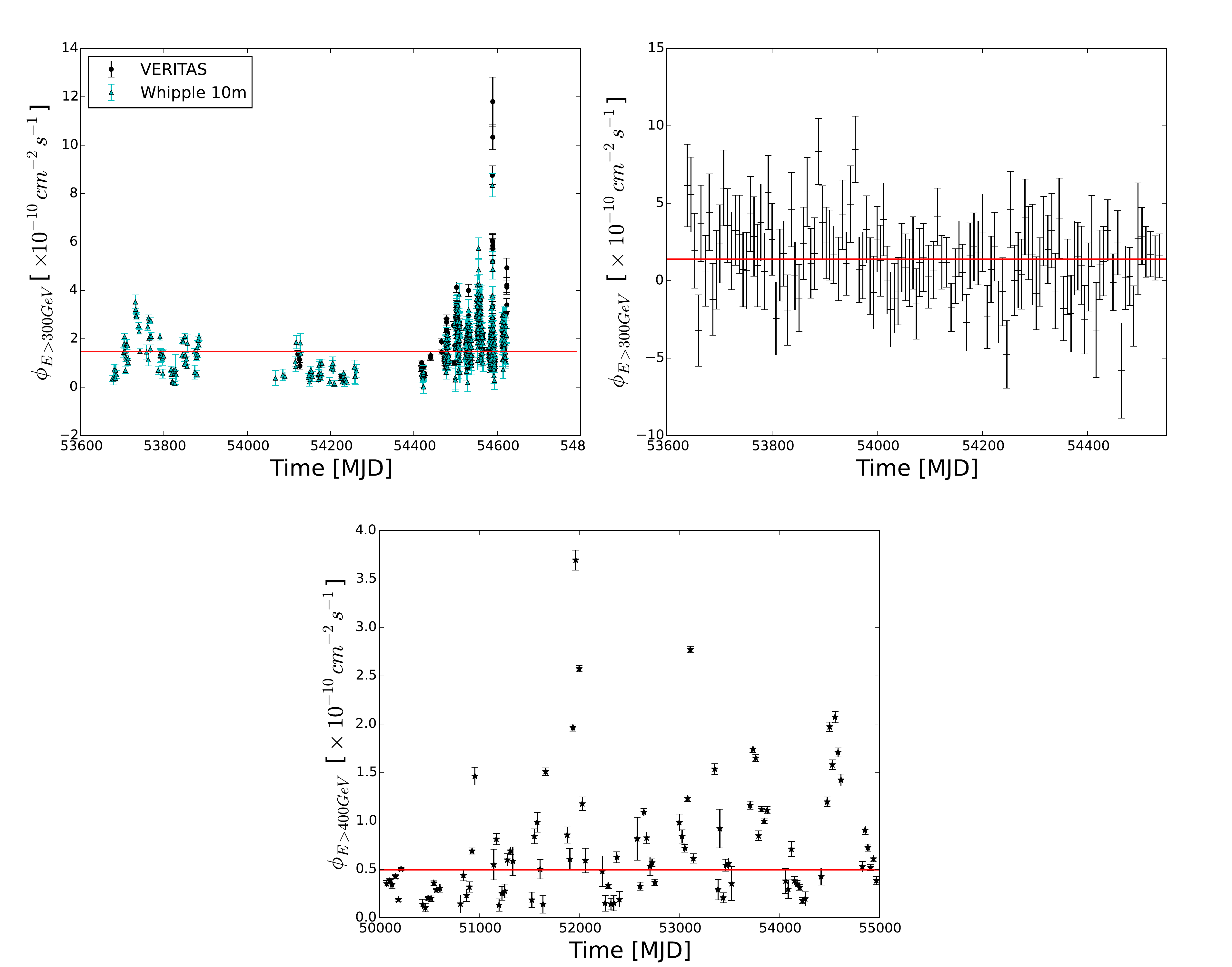}
\caption{Mrk 421 light curves from the long campaigns measured by TeV $\gamma$-ray experiments. Left-hand panel above: The data consists of 47.3 hr of VERITAS and 96 hr of  Whipple 10m telescope ($>$ 300 GeV) acquired between 2006 January and 2008 June. Right-hand panel above:  The data consists of  3 years (between 2005 September 21 and 2008 March 15) of observation with Milagro experiment   ($>$ 300 GeV). Panel below: The data consists of 14 years (878.4 h) of observation with Whipple 10m telescope ($>$ 400 GeV).  \label{fluxes}}
\end{figure*}

\begin{figure*}
\centering
\includegraphics[width=1.1\textwidth]{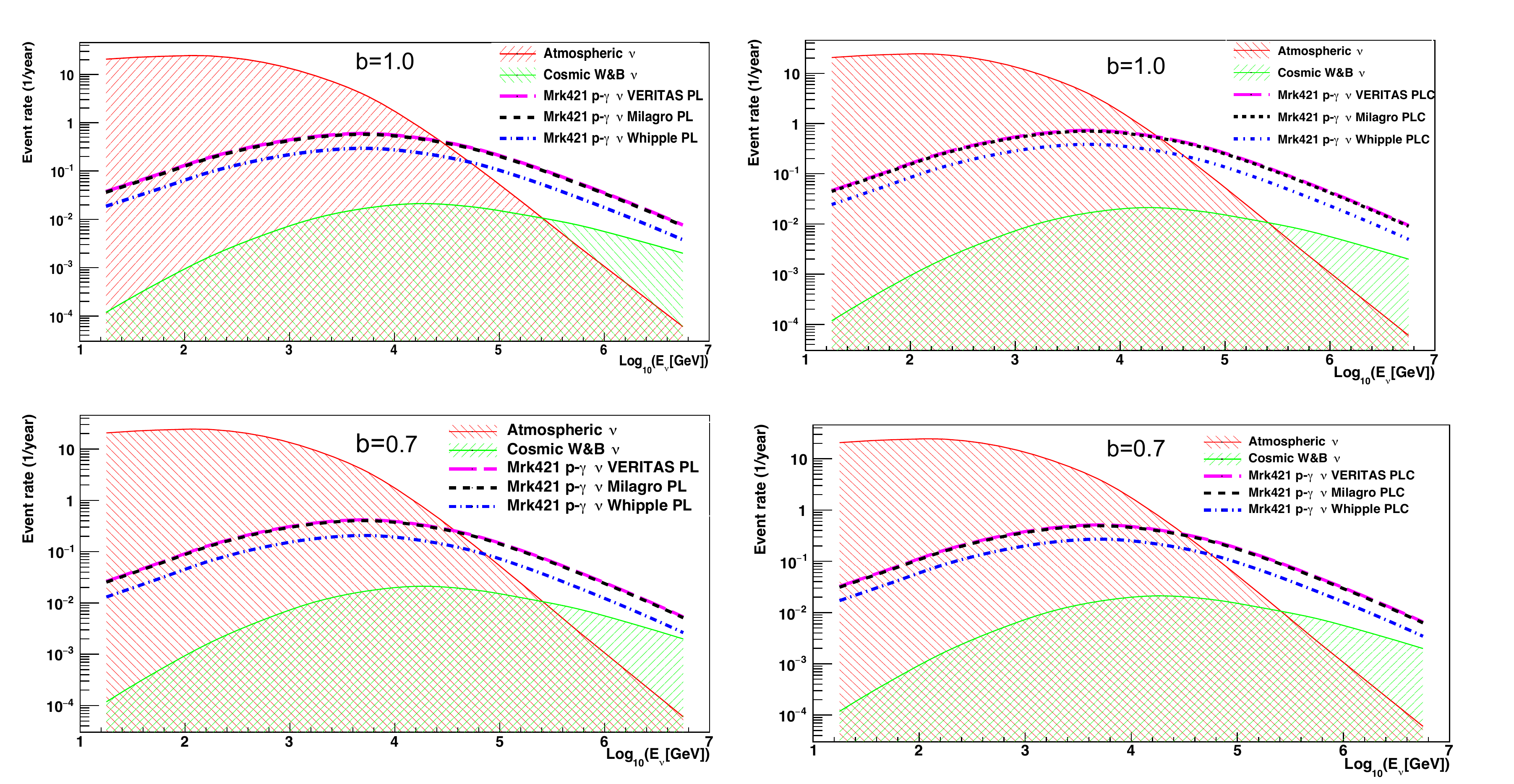}
\caption{Mrk 421 neutrino event rate is plotted as a function of neutrino energy. We report the neutrino event rate per year for signal and background, for a hypothetical km$^3$ neutrino telescope in the south hemisphere.  We normalize the neutrino event rate with the flux normalizations of the TeV $\gamma$-ray observations of Mrk 421 when a simple power law (left panel) and a power law with exponential cutoff (right panel)  are considered. The long TeV campaigns, the atmospheric and the cosmic neutrino "background" are given in the plot labels. The parameter $b$ is associated to the flux generated by  p$\gamma$ interactions. \label{neut_event}}
\end{figure*}

\begin{figure*}
\centering
\includegraphics[width=\textwidth]{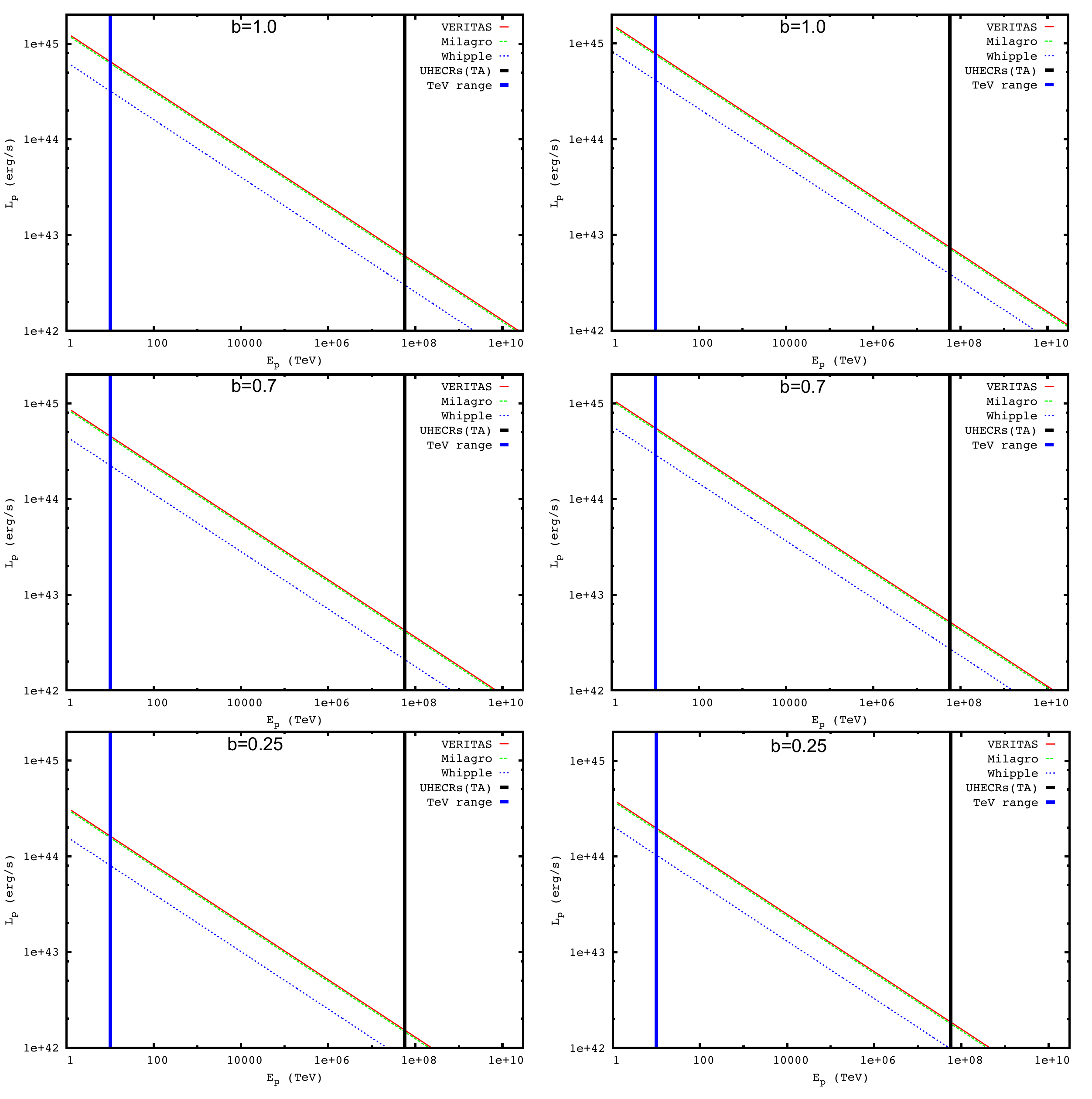}
\caption{Proton luminosity as a function of its energy is plotted.   The values of proton luminosity are normalized with the TeV $\gamma$-ray observations of Mrk 421.   In this figure we highlight two zones;  the blue solid line  represents those protons responsible for the TeV $\gamma$ production around E$_{\nu,{\rm pk}}\simeq$ 10 TeV and the black solid line represents those with energy collected in TA experiment. We consider the values of photon flux normalization fitted with a simple power law (left) and a power law with exponential cutoff (right) for b=1.0 (top), b-=0.7 (middle) and b=0.25 (bottom). \label{luminosty}}
\end{figure*}

We have introduced a hadronic model based on p$\gamma$ interactions to describe the TeV $\gamma$-ray fluxes observed from Mrk 421.   Thanks to the data collected by VERITAS telescope \citep{2011ApJ...738...25A}, Milagro observatory \citep{2014ApJ...782..110A} and Whipple 10m telescope \citep{2014APh....54....1A}, we plot the light curves of Mrk 421 and fit the integrated flux over the whole time period, as shown in fig. \ref{fluxes}.   Using the method of Chi-square $ \chi^2$ minimization as implemented in the ROOT software package \citep{1997NIMPA.389...81B}, we get the integrated flux (N$_\gamma$) as reported in table 1.  In this table we briefly summarize the main features of the long TeV campaigns. 
\begin{center}\renewcommand{\arraystretch}{1.2}\addtolength{\tabcolsep}{4.9pt}
\begin{tabular}{ l c c c c c}
 \hline \hline
 {\bf Experiments} 	&{\bf Campaings} 	& {\bf T$_{live}$}				& {\bf N$_\gamma$} \\  
 				     	&\scriptsize{(Duration)}  	&								& \scriptsize{ $(\times \,10^{-11}\,{\rm cm^{-2}\,s^{-1}})$}	   \\
\hline
\hline
\scriptsize{$^{[1]}$VERITAS}  	 &\scriptsize{ 2006 - 2008} &   \scriptsize{$\simeq$ 143.3 hr}					& \scriptsize{$1.459\pm0.008$}				                                                \\
\scriptsize{($>$ 300 GeV)}			 &					&										         &									                     \\
\hline
\scriptsize{$^{[2]}$Milagro}  		&\scriptsize{ 2005 - 2008} &   \scriptsize{$\simeq$  900 d}				& \scriptsize{$1.401\pm0.012$}				                                                                                  \\
\scriptsize{($>$ 300 GeV)}					 &					    				&																				                \\
\hline
\scriptsize{$^{[3]}$Whipple}   	 &\scriptsize{1995 - 2009} &  \scriptsize{$\simeq$ 878.4 hr}				& \scriptsize{$0.495\pm0.009$}				                                                                                    \\
\scriptsize{($>$ 400 GeV)}					&									&																                                 \\
\hline
\end{tabular}
\end{center}
\begin{center}
\scriptsize{\textbf{Table 1. The main features of the TeV long campaigns on Mrk 421 from 1995 to 2009. The integrated fluxes were obtained after fitting the light curves over the whole time period of observations (see fig. \ref{fluxes})}.}
\end{center}
\paragraph{$^{[1]}$VERITAS.} VERITAS is made of four 12 m diameter imaging atmospheric Cherenkov telescopes and is located at  the base of the Fred Lawrence Whipple Observatory (FLWO) in south Arizona at an altitude of 1280 m. It detects $\gamma$-rays in the energy range from $\sim$ 100 GeV to $\sim$ 30 TeV. More details about VERITAS, the data calibration, and the analysis techniques can be found in \citet{2008ApJ...679.1427A}. \citet{2011ApJ...738...25A} reported observations in different states of TeV $\gamma$-ray  activity between 2006 January and 2008 June. These observations for energies above 300 GeV were performed  with VERITAS observatory and the Whipple 10 m  telescope,  47.3 hr of VERITAS  and 96 hr of Whipple.
\paragraph{$^{[2]}$Milagro.} Milagro experiment, the predecessor of High Altitude Water Cherenkov (HAWC) observatory \citep{2013APh....50...26A}, was a large water-Cherenkov detector situated at  the Jemez Mountains near Los Alamos, New Mexico (USA) at an altitude of 2630 m. The main detector consisted of a central 80 m $\times$ 60 m $\times$ 8 m water reservoir with 723 photomultiplier tubes (PMT) arranged in two layers (for details see \cite{2004ApJ...608..680A} and \cite{2008ApJ...688.1078A}).  It was designed to detect TeV $\gamma$-rays at energies between 100 GeV and 100 TeV. Mrk 421 was continually observed by Milagro for a period of 906 days (between  2005 September 21 and 2008 March 15) with a significance of 7.1 standard deviations.  The observed spectrum  for energies above 300 GeV was consistent with previous observations performed by VERITAS (spectral index equal to 2.3 and an exponential energy cutoff between 2.2 and 5.6 TeV)  \citep{2014ApJ...782..110A}.
\paragraph{$^{[3]}$Whipple.} The Whipple 10m gamma-ray telescope was located at the Fred Lawrence Whipple Observatory (FLWO) in southern Arizona, USA. The predecessor of VERITAS, this telescope was in operation since 1968 and detected the first TeV gamma-ray source, the Crab Nebula in 1989 \citep{1996SSRv...75....1W}.  This Telescope was sensitive in the energy range from 200 GeV to 20 TeV.  More details about the descriptions of Whipple observing modes, analysis procedures and the GRANITE-III camera  can be found in \citep{1991AIPC..220..321P, 1993ApJ...404..206R, 2007APh....28..182K}.  Mrk 421 was studied over a 14-year time period (878.4 h, between 1995 December and May 2009)  with the Whipple 10m telescope. For energies above 400 GeV,  the time-average flux reported during this period was $0.446 \pm 0.008$ Crab flux units.\\
\\
By considering the target photon flux $(\nu F_\nu)_{0.5\,{\rm keV}} \simeq 5\times 10^{-10}\, {\rm erg\,cm^{-2}\, s^{-1}}$  \citep{2011ApJ...736..131A} and from eq. (\ref{den}), it is possible to obtain a photon density equal to $n_{\gamma}\simeq 22.38\times 10^{12}\, {\rm cm}^{-3}$. As one can see the amount of photons is copious, therefore  p$\gamma$ interactions become an effective process. Assuming the values reported by \cite{2011ApJ...736..131A}: the emitting region $r_d=4\times 10^{14}\,{\rm cm}$, the  bulk Lorentz factor $\Gamma=10$, the magnetic field $B=50$ Gauss and the break photo-pion energy {\small $\epsilon_{\pi^0,\gamma,c}\simeq 250\, {\rm GeV}$}, we compute, through p$\gamma$ interactions, the neutrino flux (eq. \ref{Av_Ag}), the number of UHECRs (eq. \ref{nUHE1}) and the proton luminosity (eq. \ref{lum}) associated to the TeV $\gamma$-ray fluxes reported by VERITAS, Milagro and Whipple experiments. Considering a simple power law {\small $\frac{dN_\gamma}{d\epsilon_\gamma}=A_{\gamma,{\rm PL}} \left(\epsilon_\gamma/\epsilon_0\right)^{-\alpha}$} and a power law with exponential cutoff {\small $\frac{dN_\gamma}{d\epsilon_\gamma}=A_{\gamma,{\rm CPL}} \left(\epsilon_\gamma/\epsilon_0\right)^{-\alpha}\, \exp(-\epsilon_\gamma/\epsilon_{c})$},  we get the values of flux normalizations (see table 2) using the integrated fluxes reported in table 1. The values of A$_{\gamma,{\rm PL}}$ and A$_{\gamma,{\rm CPL}}$ are computed with the normalization energy $\epsilon_0= {\rm E_0}= 1\, {\rm TeV}$, the cutoff energy $\epsilon_c=4\,  {\rm TeV}$ and  spectral index $\alpha=$ 2.3 \citep{2014ApJ...782..110A, 2011ApJ...738...25A}.
\begin{center}\renewcommand{\arraystretch}{1.2}\addtolength{\tabcolsep}{6pt}
\begin{tabular}{| l c  c  c |}
 \hline \hline

{\bf Experiments} 	&{\bf E$_{th}$} 	&{\bf A$_{\gamma,{\rm PL}}$} & {\bf A$_{\gamma,{\rm CPL}}$}	 \\
 				     	&\scriptsize{(TeV)}  	&\scriptsize{$(\times \,10^{-11}\,{\rm TeV^{-1}\,cm^{-2}\,s^{-1}})$} &\scriptsize{$(\times \,10^{-11}\,{\rm TeV^{-1}\,cm^{-2}\,s^{-1}})$} \\
\hline
\hline

\scriptsize{VERITAS}  	 &\scriptsize{ 0.3} &\scriptsize{$3.966\pm0.022$}	&\scriptsize{$4.831\pm0.027$}		   \\
					 &					&				&	      \\
\hline

\scriptsize{Milagro}  		&\scriptsize{ 0.3} & \scriptsize{$3.805\pm0.033$}	& \scriptsize{$4.636\pm0.040$}			\\
					 &					&				&	    \\
\hline
\scriptsize{Whipple}  	 &\scriptsize{0.4} &\scriptsize{$1.955\pm0.036$}	&\scriptsize{$2.526\pm0.046$}				 \\
					 &					& 			&			   \\
\hline

\end{tabular}
\end{center}
\begin{center}
\scriptsize{\textbf{Table 2.  Values of  the flux normalizations when we consider a  simple power law {\rm A$_{\gamma,{\rm PL}}=N_\gamma/[\int_{\epsilon_{th}} \left(\epsilon_\gamma/\epsilon_0\right)^{-\alpha}\,d\epsilon_\gamma]$} and a power law with exponential cutoff  {\rm A$_{\gamma,{\rm CPL}}=N_\gamma/[\int_{\epsilon_{th}} \left(\epsilon_\gamma/\epsilon_0\right)^{-\alpha}\, \exp(-\epsilon_\gamma/\epsilon_{c})\,d\epsilon_\gamma]$}.  These quantities are calculated from the integrated fluxes (see table 1) and for  $\epsilon_0=1\, {\rm TeV}$, $\epsilon_c=4\,  {\rm TeV}$ and  $\alpha=$ 2.3.}}
\end{center}
By assuming that HE neutrinos and TeV $\gamma$-rays are produced by p$\gamma$ interactions, we use a Monte Carlo simulation to calculate the expected neutrino events per year in a hypothetical  km$^{3}$  Cherenkov telescope located at the southern hemisphere.  This simulation was done with the values of the spectral index, the normalization energy, the flux normalizations  of the long TeV $\gamma$-ray campaigns (see table 2)  and the bin of one square degree around Mrk 421 position.  Figure \ref{neut_event} shows the neutrino event rate per year for signal and background expected from TeV $\gamma$-ray campaigns, the atmospheric and cosmic diffuse neutrino ``backgrounds''.  Left panels in fig. \ref{neut_event} show the neutrino event rate per year when the TeV $\gamma$-ray fluxes are fitted with a simple power law and right panels when the TeV  $\gamma$-ray  fluxes  are fitted with a  power law with exponential cutoff  for b=1.0 (top panels) and 0.7 (bottom panels).  From this figure one can see that the neutrino signal begins to be higher than atmospheric neutrinos for neutrino energies above $\sim$ 30 TeV. Then, we compute this excess of $\nu_\mu$ signal above the "background" for  b=1.0 and 0.7, as shown in table 3.  This table shows that the neutrino events predicted in three years are 1.02, 0.96 and 0.30  (1.41, 1.35 and 0.51) when we consider b=1.0 and TeV $\gamma$-ray flux normalizations fitted with a simple power law (power law with exponential cutoff). Similarly, the neutrino events are 0.72, 0.66 and 0.21 (0.99, 0.96 and 0.30) when we take into account  b=0.7 for the same TeV $\gamma$-ray flux normalizations.  Taking into consideration the fact that no neutrino track events in IceCube experiment were associated to the blazar Mrk 421 during three consecutive years of data taking, we have chosen the values of b=1.0  and b=0.7 as upper limits  when the long TeV $\gamma$-ray observations have been fitted with a simple power law and a power law with exponential cutoff, respectively.  We obtain that these long observations (fitted with a simple power law) could be entirely associated to the p$\gamma$ interactions for the case of a simple power law fitting, whereas less than ~70\% (b$\leq$0.7) could be only associated to these interactions when a power law with exponential cutoff is used.
\begin{center}\renewcommand{\arraystretch}{1}\addtolength{\tabcolsep}{16pt}
\begin{tabular}{| l c   | l}
 \hline \hline

{\bf Experiments} 	&{\bf  Neutrino events per year} 	 \\
 				        &	{\bf \scriptsize{b=1.0} \hspace{2.5cm} \scriptsize{b=0.7}}	 \\
 				     	& \scriptsize{${\bf [a]}\hspace{1.2cm}\,{\bf [b]} \hspace{1.5cm}  {\bf [a]}\hspace{1.2cm}\,{\bf [b}]$      } 	 \\

\hline
\hline

\scriptsize{VERITAS}  	& \scriptsize{$0.34\hspace{1.0cm}\,0.47 \hspace{1.5cm} 0.24\hspace{1.0cm}\,0.33 $   }\\
					 &					      \\
\hline

\scriptsize{Milagro}  		 &\scriptsize{$0.32\hspace{1.0cm}\,0.45 \hspace{1.5cm}  0.22\hspace{1.0cm}\,0.32$}	\\
					 &					    \\
\hline
\scriptsize{Whipple}  	         &\scriptsize{$0.10\hspace{1.0cm}\,0.17 \hspace{1.5cm}  0.07\hspace{1.0cm}\,0.10$}		 \\
					 &										   \\
\hline

\end{tabular}
\end{center}
\begin{center}
\scriptsize{\textbf{Table 3.  Excess of $\nu_\mu$ signal above the background considering a hypothetical km$^3$ neutrino telescope in the southern hemisphere. The labels [a] and [b] correspond to the values obtained using the TeV  flux normalizations fitted with a simple power law and with a power law with exponential cutoff, respectively.}}
\end{center}
Due to extragalactic (eq. \ref{thet_EG}) and galactic (eq. \ref{thet_G}) magnetic fields, UHECRs are deflected between the true direction to the source, and the observed arrival direction. Regarding these considerations, the total deflection angle may be  as large as the mean value of $<\theta_T>\simeq$ 15$^\circ $\citep{2010ApJ...710.1422R}. Therefore, it is reasonable to associate at least one UHECR within a $5^\circ$  (eqs. \ref{thet_EG}) centered at Mrk 421 (see fig. \ref{fit}). By assuming that the BH jet has the potential to accelerate particles up to UHEs and from eq. (\ref{sregion}), we can see that  protons  at the emitting region can achieve a maximum energy of 2.37$\times 10^{20}$ eV. Then, considering  that proton spectrum extends up to this maximum energy \citep{2009NJPh...11f5017K, 2009NJPh...11f5016D, 2003APh....19..559P} and taking into account that  23\%  of proton fraction will survive over a distance of 134 Mpc \citep{2011ARA&A..49..119K},  we plot the proton luminosity  normalized with the TeV fluxes fitted with a simple power law (left panels) and with a power law with exponential cutoff (right panels) for b=1.0 (top panels), 0.7 (middle panels) and 0.25 (bottom panels), as shown in Figure \ref{luminosty}. This figure also shows  two proton energies marked in blue and black colors. The blue solid line represents  those protons at 10 TeV  energies while the black solid line those at 57 EeV. One can see that proton luminosity is a decreasing function of proton energy and an increasing function of the parameter b.   For instance,  considering b=1.0 and the proton spectrum normalized with TeV fluxes fitted with a simple power law, the proton luminosity required to describe the TeV $\gamma$-ray fluxes lies in the range $3.20\times 10^{44}\, {\rm erg/s} \leq L_p \leq 6.49\times 10^{44}\, {\rm erg/s}$ and  the UHE proton luminosity (at $57\times 10^{18}$ eV)  lies in the range $3.01\times 10^{42}\, {\rm erg/s} \leq L_p \leq 6.09\times 10^{42}\, {\rm erg/s}$. Similarly, considering the TeV flux normalization fitted with a power law with exponential cutoff (right panels), the proton luminosity at 10 TeV lies in the range $4.14\times 10^{44}\, {\rm erg/s} \leq L_p \leq 7.92\times 10^{44}\, {\rm erg/s}$ and  the UHE proton luminosity in the range $3.88\times 10^{42}\, {\rm erg/s} \leq L_p \leq 7.42\times 10^{42}\, {\rm erg/s}$.  Additionally,  from eq. (\ref{nUHE1}) we estimate the number of events expected in TA experiment for b=1.0, 0.7 and 0.25.  Table 4 shows that the number of events  lie in the range  $1.661\leq N_{\tiny UHECR}\leq 3.369$ ($2.146\leq N_{\tiny UHECR}\leq 4.104$), $1.163\leq N_{\tiny UHECR}\leq 2.358$ ($1.502\leq N_{\tiny UHECR}\leq 2.873$) and $0.415\leq N_{\tiny UHECR}\leq 0.842$ ($0.537\leq N_{\tiny UHECR}\leq 1.026$) for  b=1.0, 0.7 and 0.25 when proton spectrum is normalized with TeV fluxes fitted with a simple power law (power law with exponential cutoff), respectively.  One can see that the number of predicted events is higher than one for b $\geq$ 0.3 (0.25) and the proton spectrum normalized with TeV fluxes fitted with a simple power law (power law with exponential cutoff).
\begin{center}\renewcommand{\arraystretch}{1.2}\addtolength{\tabcolsep}{5pt}
\begin{tabular}{| l c  c  |}
 \hline \hline

{\bf Experiments}	& {\bf Campaigns} 	& {\bf N$_{\tiny UHECR}$}	 \\
 				     	&\scriptsize{ (Duration) } 	&{\bf \scriptsize{b=1.0} \hspace{3.9cm} \scriptsize{b=0.7} \hspace{3.9cm} \scriptsize{b=0.25}} \\
 				     	&				 	& \scriptsize{${\bf [a]}\hspace{1.4cm}\,{\bf [b]}\hspace{2.4cm} {\bf [a]}\hspace{1.6cm}\,{\bf [b]}\hspace{2.4cm} {\bf [a]}\hspace{1.5cm}\,{\bf [b]}$} 	 \\

\hline
\hline

\scriptsize{VERITAS}  	 &\scriptsize{ 2006 - 2008} &\scriptsize{$3.369\pm0.019\hspace{0.3cm}\,4.104\pm0.023 \hspace{0.8cm} 2.358\pm0.013\hspace{0.5cm}\,2.873\pm0.016\hspace{0.8cm} 0.842\pm0.005\hspace{0.5cm}\,1.026\pm0.006$}\\
					 &					&					      \\
\hline

\scriptsize{Milagro}  		&\scriptsize{ 2005 - 2008} &\scriptsize{$3.233\pm0.028\hspace{0.3cm}\,3.939\pm0.034\hspace{0.8cm}   2.263\pm0.020\hspace{0.5cm}\,2.758\pm0.024\hspace{0.8cm}   0.808\pm0.007\hspace{0.5cm}\,0.985\pm0.009$}	\\
					 &					&					    \\
\hline
\scriptsize{Whipple}  	 &\scriptsize{1995 - 2009} &\scriptsize{$1.661\pm0.031\hspace{0.3cm}\,2.146\pm0.039 \hspace{0.8cm}1.163\pm0.022\hspace{0.5cm}\,1.502\pm0.027\hspace{0.8cm}0.415\pm0.008\hspace{0.5cm}\,0.537\pm0.010$}		 \\
					 &					& 						   \\
\hline

\end{tabular}
\end{center}
\begin{center}
\scriptsize{\textbf{Table 4.  Number of UHE protons to be expected in TA experiment. These estimations were computed through the p$\gamma$ interactions and normalized with the TeV $\gamma$-ray flux fitted with a simple power law ([a]) and with a power law with exponential cutoff ([b]).}}
\end{center}
%
%
\section{Results and Conclusions}
We have introduced a hadronic model through p$\gamma$ interactions to describe the long TeV $\gamma$-ray campaigns of the BL Lac Mrk 421 and to correlate them with the expected HE neutrino and  UHECR fluxes around Mrk 421.  First of all, we have obtained the integrated fluxes over the whole time period of  long TeV observations performed by  VERITAS, Milagro and Whipple experiments (fig. \ref{fluxes}).  Considering a simple power law and a power law with exponential cutoff we have computed the flux normalizations for $\epsilon_0=1\, {\rm TeV}$, $\epsilon_c=4\,  {\rm TeV}$ and  $\alpha=$ 2.3 \citep{2014ApJ...782..110A, 2011ApJ...738...25A}, as shown in table 2.\\ 
Correlating the TeV $\gamma$-ray and HE neutrino fluxes through p$\gamma$ interactions,  we have plotted (through MC simulation) the neutrino event rate per year expected from the TeV $\gamma$-ray campaigns, the atmospheric and the cosmic diffuse neutrino ``backgrounds'', assuming a hypothetical km$^3$ Cherenkov telescope in the southern hemisphere.  Considering the neutrino energies $E_\nu\geq$ 30 TeV for which the signal is higher than atmospheric neutrinos, we estimated  the neutrino events  expected in a hypothetical km$^3$, as shown in table 3.  Taking into consideration the neutrino observations reported by The IceCube Collaboration during three consecutive years  \citep{2014arXiv1405.5303A},   we found that the long TeV $\gamma$-ray flux is consistent with production by p$\gamma$ interactions when these observations are fitted with a simple power law and  less than ~70\% (b$\leq$0.7) when they are fitted with a power law with exponential cutoff.\\
Taking into account the maximum energy  achieved by Fermi-accelerated protons at the emitting region ($2.7\times 10^{20}$ eV) and the TA  exposure, we normalized the proton spectrum with the long TeV observations (eq. \ref{Apg}) in order to estimate the number of UHECRs above 57 EeV,  as shown in  table 4. Assuming that at least one UHECR was detected by TA experiment around the Mrk 421 position, we found that  more than ~30\% (b$\geq$0.3) of the long TeV $\gamma$-ray observations could be associated to the p$\gamma$ interactions when these observations are fitted with a simple power law and more than ~25\% (b$\geq$0.25) when they are fitted with a power law with exponential cutoff.\\
After comparing the TeV $\gamma$-ray fluxes collected from Mrk 421 with the neutrino and UHECR fluxes through p$\gamma$ interactions, we conclude  that only from $\sim$ 25\% to 70\% of TeV $\gamma$-ray fluxes described with a power law  with exponential cutoff can come from the p$\gamma$  interactions and more than $30$\% when these observations are described with a simple power law.  It is worth noting that the increasing statistics of IceCube and TA experiments will give more evidence of the presented scenario. \\
Considering p$\gamma$ interactions to correlate the TeV $\gamma$-ray fluxes  with the neutrino and UHECRs reported by  IceCube and Telescope Array experiments, a lepto-hadronic model might be consistent with the observations, but it is not required.\\ 
\section*{Acknowledgements}
We thank the anonymous referee  for a critical reading of the paper and valuable suggestions that helped improve the quality and clarity of this work.  We also thank to Francis Halzen, Ignacio Taboada, Teresa Montarulli, Patrick Moriarty and William Lee for useful discussions, Matthias Beilicke, Patrick Moriarty and Barbara Patricelli for sharing with us the data. Also we thank to TOPCAT team for the useful sky-map tools.  This work was supported by Luc Binette scholarship, Galilei postdoctoral grant and the projects IG100414.
%
%
%

%
%
\clearpage
\clearpage
\clearpage
\end{document}